\documentclass[10pt]{iopart}
\usepackage{graphicx}
\begin{document}

\title[Strained band edge characteristics: GaAs, GaSb, InAs and InSb]{Strained band edge characteristics from hybrid density functional theory and empirical pseudopotentials: GaAs, GaSb, InAs and InSb}

\author{Asl{\i} \c{C}akan$^1$, Cem Sevik$^2$ and Ceyhun Bulutay$^1$}

\address{$^1$ Department of Physics, Bilkent University, Bilkent, Ankara 06800, Turkey}
\address{$^2$  Faculty of Engineering, Anadolu University, Eski\c{s}ehir 26555, Turkey}

\ead{bulutay@fen.bilkent.edu.tr}
\vspace{10pt}

\begin{abstract}
The properties of a semiconductor get drastically modified when the crystal point group symmetry is broken
under an arbitrary strain. We investigate the family of semiconductors consisting of GaAs, GaSb, InAs and InSb, 
considering their electronic band structure and deformation potentials subject to various strains
based on hybrid density functional theory. 
Guided by these first-principles results, we develop strain-compliant local pseudopotentials
for use in the empirical pseudopotential method (EPM). We demonstrate that the newly proposed empirical 
pseudopotentials perform well close to band edges and under anisotropic crystal deformations. 
Using EPM, we explore the heavy hole-light hole mixing characteristics under different stress directions 
which may be useful in manipulating their transport properties and optical selection rules. 
The very low 5~Ry cutoff targeted in the generated pseudopotentials paves the way for large-scale EPM-based 
electronic structure computations involving these lattice mismatched constituents.
\end{abstract}





%

\section{Introduction}
Among group-III-V compound semiconductors crystallizing in the cubic phase, the family of GaAs, GaSb, InAs and InSb span the widest direct bandgap range from 1.51~eV down to 0.235~eV at zero temperature
\cite{Meyer2001}. Their quantum dots such as InAs/GaAs, InSb/GaSb and other combinations have been in
the spotlight due to their applications as light emitters, and because of their potential for emerging quantum 
information technologies \cite{InSb1999,he2004,zielinski13,huo2014}. For this lattice mismatched family the 
inevitable overarching theme in their heterostructures is strain. As gradually 
appreciated over the decades, strain has been a game changer for materials and especially for semiconductors. 
As a matter of fact much of the novel features in these quantum dots, and the very existance of self-assembly are owed to strain \cite{stangl04}.
This is also the case for bulk properties. For instance, the nature of bandgap changes from indirect to direct under tensile strain in the case of germanium \cite{Ge-mit}, or higher hole mobilities are attained under strain
in group-III-V arsenides, and group-III-V antimonides \cite{nainai}.

Historically, within the context of strain the key concept of deformation potentials were introduced as early as 1950's by the pioneers of semiconductor physics, Bardeen and Shockley~\cite{Bardeen50} for uniformly strained silicon and germanium, consequently generalized to multivalley case by Herring and Vogt~\cite{Herring57}.
The shear deformation parameters related to the conduction band edge for Si has been worked out by Sham in the framework of a pseudopotential rigid ion model~\cite{Sham1963}. In the 1960's, both conduction and valence band deformation potentials of Si under hydrostatic and uniaxial stress have been studied using self-consistent perturbation theory by Kleinman~\cite{Kleinman1963}. Subsequently, Ge and GaAs crystals under uniaxial stress have been investigated to define splitting of valence band and the direct bandgap energy shift within optical reflection measurements by Balslev~\cite{Balslev1967}. Likewise, the effects of uniaxial stress along [001], [110] and [111] on the electronic structure of Ge, GaAs and Si were investigated by Pollak and Cardona with the calculation of hydrostatic and shear deformation potentials for both conduction and valence bands~\cite{Cardona68}. 
The decade of 1970's was stagnant except Pollak's review \cite{Pollak1973}, and the seminal monograph of Bir and Pikus on strain-induced effects in semiconductors~\cite{BirPikus}.
A revival started with Landolt-B\"{o}rnstein experimental data compilation for III-V semiconductors in 1980's~\cite{Landolt1982}. Afterwards, \emph{ab initio} theoretical studies soared \cite{Walle1989-1, Walle1989-2, Wei-Zunger1999, Li2006-2, Li2006-1, Tawinan2011, Zielinski2012, Bester2012}. As these \emph{ab initio} calculations suffer from well-known bandgap errors and require high computational cost \cite{Martin2004}, semiempirical methods have also been preferred \cite{Cardona-Blacha1984,Cardona1987,Allan1988,OReilly1992,Williamson2000,Niquet2006,FischettiJAP2010,He2012}. Particularly, O'Reilly using tight-binding method has taken into account the crystals under biaxial compression and tension to study [001] axial deformation potential $b$ in group-III-V semiconductors~\cite{O'Reilly1986}. He showed that in biaxially strained materials, the heavy hole band could show light hole type characteristics \cite{OReilly1992}, which is recently reinstated \cite{FischettiJAP2010,Bester2012}. Nowadays, this result has also ramifications in quantum information science and technology.

To meet the demand for a more up-to-date and reliable semiconductor data Vurgaftman, Meyer, and Ram-Mohan (VMR) carried out the most recent compilation, still more than a decade ago \cite{Meyer2001}. Inevitably, they expressed the deformation potentials as ranges and gave recommendations which are for some materials ambiguous. Notably, there are experimental discrepancies on $b$ and $d$ biaxial deformation potentials. 
Therefore, the necessity for reliable deformation potentials for common  III-V semiconductors is still a pressing issue within the community.

Over the last decade, density functional theory (DFT) calculations with hybrid functionals have received increasing attention as they offer a remedy for the well-known local density approximation (LDA) failures \cite{Martin2004} with the approach by Heyd-Scuseria-Ernzerhof \cite{HSE2003}. Hence, it is a well-suited reliable method for studies on strained materials. As a matter of fact, using hybrid functionals Van de Walle group has calculated shear deformation potentials of GaN and InN to explore the effect of strain in polarization switching in InGaN/GaN quantum wells \cite{Walle2010}, and observed pronounced nonlinear dependence on strain for AlN, GaN, InN and ZnO \cite{Walle2011}. For the wurtzite phases of InAs and InP, Hajlaoui \emph{et al} have obtained the deformation potentials and revealed the failure of the quasi-cubic approximation \cite{Hajlaoui2013}.
Interestingly, to the best of our knowledge a detailed hybrid DFT study for the strained cubic phase GaAs, GaSb, InAs and InSb has not been undertaken.

Our first aim with this work is to extract reliable deformation potentials using the state-of-the-art hybrid DFT computations so as to 
alleviate the ambiguity in the VMR data \cite{Meyer2001} over the compounds GaAs, GaSb, InAs and InSb. Moreover,
in the light of hybrid DFT computations we develop a new set of empirical pseudopotential parameters which can reproduce the band edge hybrid DFT results under various strain profiles, while excluding the nonlocal parts and using a very low 5~Ry energy cutoff
in the interest of reduced computational budget for million-atom structures \cite{wang99a}. 
Finally, we demonstrate performance of this
empirical pseudopotential parameter set through examining the directional variation of valence band effective masses under a
uniaxial stress and shed light on the intricate heavy and light hole mixing characteristics in this material family.

\begin{table}[h!]
\caption{Comparison of our hybrid DFT and EPM unstrained  band structure values with VMR data~\cite{Meyer2001} at 0~K. 
$E_{\mbox{\tiny {gap}}}$ is the direct energy gap (eV), $\Delta_{0}$ represents valence band spin-orbit splitting (eV), and $a_0$ 
is the equilibrium lattice constant ({\AA}).}
\begin{center}
\begin{tabular}{ l c c c c c c c c}
  \hline
  \hline
      &    &\multicolumn{2}{c}{This Work}& \multicolumn{1}{c}{VMR~\cite{Meyer2001}} \\
 \cline{3-4}
 \cline{6-7}
Material	&   				&Hybrid DFT	&EPM	&                    &\\
\hline
 		&$E_{\mbox{\tiny {gap}}}$  		&1.36 	&1.51	&1.52 \\
GaAs  	&$\Delta_{0}$  	&0.36	&0.367	&0.32-0.36 \\
  		&$a_{0}$  		&5.626	&5.653	&5.653 \\
  \hline
		&$E_{\mbox{\tiny {gap}}}$ 		&0.81	&0.812	&0.811-0.813  \\
GaSb  	&$\Delta_{0}$	&0.72	&0.723	& 0.749-0.82   \\
  		&$a_{0}$  		&6.095	&6.0959	&6.0959 \\
  \hline
   		&$E_{\mbox{\tiny {gap}}}$		&0.34	&0.41	& 0.41-0.45   \\
InAs 	&$\Delta_{0}$  	&0.38	&0.388	& 0.37-0.41  \\
 		&$a_{0}$  		&6.043	&6.058	&6.058 \\
  \hline
 		&$E_{\mbox{\tiny {gap}}}$     	&0.27	&0.235	& 0.235   \\
InSb     	&$\Delta_{0}$  	&0.76	&0.763	& 0.8-0.9  \\
     		&$a_{0}$  		&6.457	&6.479	&6.479 \\
\hline
\hline
\end{tabular}
\end{center}
\label{Table-gaps}
\end{table}

\section{Theory}
\subsection{DFT with Hybrid Functionals}
Hybrid functionals such as Heyd-Scuseria-Ernzerhof (HSE) combine LDA or generalized gradient approximation (GGA) exchange-correlation functionals with Hartree-Fock (HF) exact exchange \cite{HSE2003}. One of the most advantageous features of HSE is to use conventional \emph{local} functional instead of long-range part of HF term, however, short-range part is switched to the nonlocal HF potential since the calculation of long-range part for localized basis set projected augmented wave (PAW) is troublesome and computationally costly. Conveniently, HSE reduces the high cost of hybrid functionals to within a factor of 2-4 higher than pure DFT
functionals, while providing much reliable energy bandgaps for semiconductors \cite{heyd-2004}.
Some of the available hybrid functionals are PBE0 \cite{PBE0-1999}, HSE03 \cite{HSE2003,HSE03-2006,heyd-2004}, HSE06 \cite{HSE06-2006}, and HSEsol \cite{HSEsolKresse}. For our work we have chosen HSEsol functional as it was reported to yield satisfying results for small gap semiconductors.

Our \emph{ab initio} calculations are performed using the Vienna Ab initio Simulation package (VASP) code \cite{vasp1,vasp2,vasp3}.
The PAW pseudopotentials from the standard distribution are incorporated in the calculations \cite{PAW-blochl,kresse-1999}. 
For electronic exchange-correlation functional, GGA in its PBEsol parametrization is used \cite{PBE1996,PBEsol1,PBEsol2}.
For the PAW pseudopotentials $d$ orbitals are taken as valence for cations, and conventional ones for anions. 
The lattice vectors and atomic coordinates are relaxed until the force on each atom is reduced to less than 0.001~eV/{\AA} 
and the total energy is iterated until changes in energy are lower than $10^{-6}$~eV. For each bulk material, $4\times4\times4$ $k$-grid 
and 450~eV cutoff energy are used while employing HSEsol hybrid functional. 
To impose the hydrostatic and uniaxial stress on the unit cell, we insert the strained lattice vectors.
Table~\ref{Table-gaps} shows that the calculated lattice constants are in good agreement with the experimental ones.

\subsection{Empirical Pseudopotential Method}
Recently, Kim and Fischetti offered local and nonlocal empirical pseudopotential parameters for a number of group-IV and
group-III-V semiconductors with 10~Ry cutoff energy~\cite{FischettiJAP2010}.
Relying on the success of the hybrid DFT electronic structures, we aim to tune the EPM parameters in Ref.~\cite{FischettiJAP2010},
under the conditions of excluding the nonlocal parts, and using a lower kinetic energy cutoff of 5~Ry to reduce the computational budget. 

For arbitrarily strained crystals, the pseudopotential parameters are needed not only at fixed wavenumbers but over a continuum.
For this purpose the cubic Hermite interpolation can be used due to its advantage of giving the means to control curve slopes 
at desired data points \cite{fritsch80}. Accordingly, the local pseudopotential is represented as
\begin{equation}
V(q)= V_{\mbox{\tiny {PCHIP}}}(q)\times\frac{1}{2}\left\{\tanh\left[\frac{a_5-q^2}{a_6}\right]+1\right\},
\label{psp-tanh}\end{equation}
where $\tanh (\cdot)$ part is introduced for a fast cutoff of the pseudopotential at large wavenumbers, $q$ involving  the $a_5$ and $a_6$ fitting 
parameters; the same values as in Ref.~\cite{FischettiJAP2010} are retained as displayed in Table~\ref{Table-epm} which
should be in Hartree atomic units, including $q$ in $\tanh$ part. 
$V_{\mbox{\tiny {PCHIP}}}(q)$ represents the piecewise cubic Hermite interpolating polynomial (PCHIP) \cite{fritsch80} of the 
symmetric and antisymmetric local form factors which consists of potentials and their slopes at certain wavenumbers. 
For the unit interval [0,1] the PCHIP has the form
\begin{eqnarray}
V_{\mbox{\tiny {PCHIP}}}(q) & = & \left( 2q^3-3q^2+1\right)V_0+\left( q^3-2q^2+q\right)s_0 \nonumber\\
& & + \left( -2q^3+3q^2\right)V_1+\left( q^3-q^2\right)s_1\, ,
\label{pchip}\end{eqnarray}
where $V_0$ and $V_1$  ($s_0$ and $s_1$) are the potential (slope) values at either end of the interval.

A general drawback of empirical pseudopotentials is that they are frozen in the sense that they lack the self-consistency loop to 
adapt for changes in the chemical environment such as bond lengths and directions as would arise under strain.
To fix this in the level of hydrostatic strains without a computational overhead, Williamson \emph{et al} introduced
an additional fitting parameter $\gamma$ so that a hydrostatic strain-dependent pseudopotential is formed as  \cite{Williamson2000}
\begin{equation}
V(q;\epsilon)=V(q)\left[ 1+\gamma\,\epsilon_H\right]\, ,
\label{strain-dependent-psp}
\end{equation}
where $\epsilon_H=\epsilon_{xx}+\epsilon_{yy}+\epsilon_{zz}$ refers to hydrostatic (volumetric) strain.

For an even better strain performance, we combine Kim and Fischetti local EPM form factors \cite{FischettiJAP2010}
with the hydrostatic strain parameter, $\gamma$  \cite{Williamson2000}. 
Then, we optimize this set including the symmetric spin-orbit coupling parameter $\lambda_S$ \cite{Williamson2000,wang99c},
choosing the target values as the experimental bandgaps, and our hybrid DFT deformation potentials, and spin-orbit splittings.
Note that in contrast to Ref. \cite{FischettiJAP2010}, we fit EPM to the zero-temperature bandgaps shown in Table~\ref{Table-gaps}.
Table~\ref{Table-epm} presents our EPM parameters at standard wave numbers of the zinc-blende structure: 0,  
$\sqrt{3}$, $\sqrt{4}$, $\sqrt{8}$, $\sqrt{11}$, all in units of $2\pi/a_0$, where $a_0$ is the unstrained lattice constant of the crystal; 
the form factors and their slopes are set to zero beyond a high value, $q^2>50$.
Regarding the associated units for these parameters, other than the $\tanh$ part 
which was separately mentioned above, $V(q)$ in (\ref{psp-tanh}) comes out in Ry when wavenumber, $q$ is used in units of $2\pi/a_0$.
$V^{s,a}(q=0)$  are adjusted to the \emph{ab initio} natural band offsets under unstrained conditions offered in Ref.~\cite{lineup09}.
It can be seen from Table~\ref{Table-effmass} that our local EPM conduction and valence band edge effective mass values are in 
reasonable agreement with Ref. \cite{FischettiJAP2010}, and the latter is in fair agreement with the VMR data \cite{Meyer2001}.

\begin{table*}[t]
\caption{Fitted local EPM parameters: form factors, and their slopes for cubic Hermite interpolation,
symmetric component of the spin-orbit coupling parameter $\lambda_S$, and hydrostatic strain parameter $\gamma$,
asymptotic cutoff parameters $a_5$, $a_6$. Refer to text for the units associated with these parameters. 
}
\begin{tabular}{@{}llccccccccccccc}
\br
  & &\multicolumn{4}{c}{Material}  \\
   \cline{3-6}
  			&Parameter	    &GaAs    &InAs	&GaSb	&InSb \\
\mr

  Local Form Factors	&$V_0^{s}$	    &-0.6421 &-0.5469 	&-0.5266 &-0.4246 \\
			&$V_{\sqrt{3}}^{s}$ &-0.2350 &-0.2070 	&-0.2043 &-0.1990 \\
                	&$V_{\sqrt{8}}^{s}$ & 0.0150 & 0.0000   & 0.0000 & 0.0115 \\
			&$V_{\sqrt{11}}^{s}$& 0.0729 & 0.0465 	& 0.0601 & 0.0334 \\
			&$V_0^{a}$	    &-0.1040 &-0.0880 	&-0.0470 &-0.0450 \\
                	&$V_{\sqrt{3}}^{a}$ & 0.0760 & 0.0540  	& 0.0330 & 0.0416 \\
                	&$V_{\sqrt{4}}^{a}$ & 0.0570 & 0.0466 	& 0.0280 & 0.0350 \\
                	&$V_{\sqrt{11}}^{a}$& 0.0061 & 0.0070  	& 0.0054 & 0.0060 \\ 
\mr
  Slopes for PCHIP 	& $s_0^{s}$		& 0.0000 & 0.0000 & 0.0000 & 0.0000   \\
			& $s_{\sqrt{3}}^{s}$	& 0.0699 &-0.1760 &-0.1668 &-0.1357   \\
                        & $s_{\sqrt{8}}^{s}$ 	& 0.1250 & 0.1250 & 0.1400 & 0.0606 \\
                        & $s_{\sqrt{11}}^{s}$	& 0.0596 &-0.0062 &-0.0819 & 0.0100   \\
                        & $s_0^{a}$		& 0.0000 & 0.0000 & 0.0000 & 0.0000 \\ 
 			& $s_{\sqrt{3}}^{a}$ 	& 0.0250 &-0.0350 &-0.0500 &-0.0500  \\
                        & $s_{\sqrt{4}}^{a}$ 	&-0.1150 &-0.0900 &-0.0400 &-0.0400  \\
                        & $s_{\sqrt{11}}^{a}$	&-0.0100 &-0.0220 &-0.0300 &-0.0300  \\
\mr
  Asymptotic parameters      & $a_{5}$ & 4.05   & 4.50   &  4.00  & 3.90  \\
                             & $a_{6}$ & 0.39   & 0.41   &  0.30  & 0.30  \\
\mr
Hydrostatic strain parameter&$\gamma$		 &-1.7392 &-0.1046 &-2.1285 &-1.4260\\
Spin-orbit coupling parameter (Ry) &$\lambda_S$  & 0.0213 & 0.0205 & 0.0385 &0.0377 \\
Cutoff energy (Ry) & $E_{\mbox{\tiny {cutoff}}}$ &5.00    &4.85    &5.00    &4.85   \\
\br
\end{tabular}
\label{Table-epm}
\end{table*}

\begin{table*}[t]
\caption{Effective masses (in free-electron mass, $m_0$) at $\Gamma$ point in \emph{k}-space for conduction band 
($m_e^{*\Gamma}$), heavy hole $m_{hh}^{*\Gamma}$, light hole $m_{lh}^{*\Gamma}$ and split-off $m_{so}^{*\Gamma}$ 
bands, compared with Kim and Fischetti (KF)~\cite{FischettiJAP2010}, and Vurgaftman, Meyer and Ram-Mohan (VMR) 
data~\cite{Meyer2001}.}
\hspace*{-1.8cm}
\begin{tabular}{ l c c c c c c c c c c c c c}
  \br
Material & &$m_e^{*\Gamma}$ & $m_{hh}^{*\Gamma}[001]$&$m_{hh}^{*\Gamma}[110]$&$m_{hh}^{*\Gamma}[111]$&
$m_{lh}^{*\Gamma}[001]$&$m_{lh}^{*\Gamma}[110]$&$m_{lh}^{*\Gamma}[111]$&$m_{so}^{*\Gamma}$ \\
\mr
       &This work&0.082&0.439&0.845&1.143&0.111&0.099&0.096&0.218\\
GaAs   &KF&0.082&0.382&0.696&0.903&0.106&0.094&0.091&0.206\\
       &VMR                  &0.063&0.388&0.658&0.920&0.089&0.081&0.079&0.33-0.388\\
  \mr
      &This work&0.054&0.363&0.724&1.027&0.066&0.060& 0.059&0.20\\
GaSb  &KF&0.049 &0.289 &0.534 &0.712&0.056&0.052&0.050&0.19\\
      &VMR&0.041 &0.23 &-- &0.57&--&0.05&--&0.14\\
  \mr
      &This work&0.030&0.433& 0.814&1.127 &0.038&0.036& 0.036&0.127 \\
InAs  &KF&0.026&0.31&0.547&0.720&0.032&0.03&0.03&0.109\\
      &VMR&0.023&0.39& 0.98&0.757& 0.042&0.041&0.014&0.09-0.15\\
  \mr
     &This work& 0.022&0.357&0.714&1.049&0.024&0.023&0.023&0.172  \\
InSb  &KF&0.017 &0.304&0.534&0.705& 0.019&0.018&0.018&0.155\\
      &VMR&0.014 &0.26& --&0.68& 0.015&0.015&--&0.19\\
\br
\end{tabular}
\label{Table-effmass}
\end{table*}

\section{Results}

\subsection{Hybrid DFT Results}
\begin{figure}
\begin{center}
\includegraphics[width=0.7\textwidth]{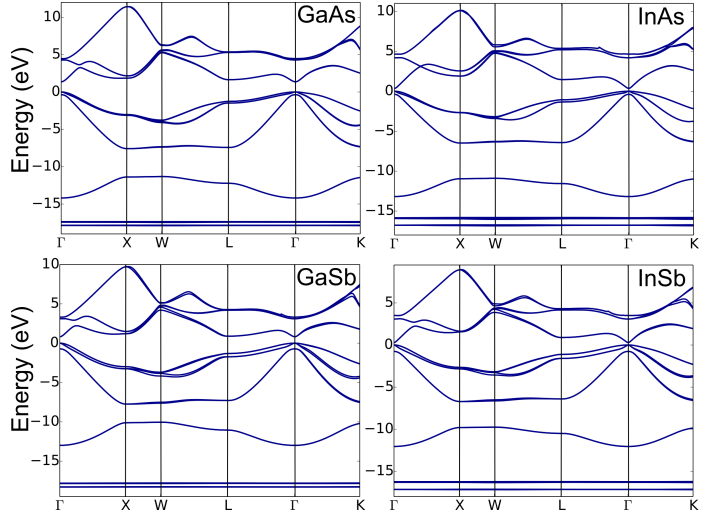}
\end{center}
\caption{Full zone band structures for unstrained GaAs, GaSb, InAs and InSb
obtained using hybrid functional DFT. Valence band maximum is set to zero energy level separately
for each material.}
\label{fig1}
\end{figure}

The hybrid functional DFT band structures  for unstrained GaAs, GaSb, InAs and InSb including the spin-orbit interaction are displayed
in Figure~\ref{fig1}. For these direct bandgap compounds, computed bandgap values are within 10\% agreement 
with the experimental values in Table~\ref{Table-gaps}. If desired, further improvement is possible by slightly adjusting the
so-called range separation parameter of the hybrid functionals \cite{moussa12}.

Next, from hybrid DFT calculations we work out the deformation potentials $a_{\mbox{\tiny {gap}}}$, $b$ and $d$ 
under the relevant stress conditions: hydrostatic, uniaxial stress along [001] and [111], respectively \cite{yu-cardona}. 
Our results are listed in Table~\ref{Table-deformation} together with VMR \cite{Meyer2001}, and other 
representative data from the literature. Notably, in some cases the VMR ranges and/or suggested values deviate 
substantially from the hybrid DFT results.

\begin{table*}[t]
\caption{Deformation potentials $a_{\mbox{\tiny {gap}}}$, $b$ and $d$ in eV units. Here, EPM values are fitted to hybrid DFT results.}
\hspace*{-0.8cm}
\begin{tabular}{ l c c c c c c c c}
  \hline
  \hline
      &  & & \multicolumn{1}{c}{This Work}& &\multicolumn{2}{c}{Literature}  \\
 \cline{3-4}
 \cline{6-7}
Material&   &&Hybrid DFT \& EPM&   & VMR$^a$ &       Others   &  \\
\hline
\quad    &$a_{\mbox{\tiny {gap}}}$&&-8.69& &[-20.4, -6.5]& -8.33$^d$, -8.76$^e$, -8.33$^h$, -8.44$^j$, -7.25$^k$ \\
GaAs     &$b$      &&-2.13& &[-3.9, -1.6] & -2.0$^{c,d}$, -1.7$^f$, -1.9$^{h,e}$, -2.79$^{b,h}$ \\
\quad    &$d$      &&-4.77& &[-6.0, -2.7] & -4.23$^{h,e}$, -4.5$^c$, -4.77$^g$, -7.5$^b$ \\
  \hline
\quad    &$a_{\mbox{\tiny {gap}}}$ &&-8.44& &-8.3  & -7.64$^h$, -7.01$^k$ \\
GaSb     &$b$       &&-2.23& &-1.6 &-1.6$^b$, -1.9$^f$,-2.0$^c$,  -2.3$^g$  \\
\quad    &$d$       &&-5.0 & &-3.98&-3.98$^g$, -4.7$^c$, -4.8$^i$, -5.0$^b$ \\
  \hline
\quad    &$a_{\mbox{\tiny {gap}}}$&&-5.95& &[-16.9, -6.08]& -6.12$^e$, -6.08$^h$, -4.93$^k$  \\
InAs     &$b$      &&-1.76& &[-5.9, -1.0] & -1.72$^b$, -1.55$^{h,e}$, -1.7$^{f,d}$, -1.8$^c$, -2.33$^g$ \\
\quad    &$d$      &&-4.25& &[-8, -2.57] & -3.3$^b$, -3.6$^c$, -3.1$^{h,e}$, -3.83$^g$ \\
  \hline
\quad       &$a_{\mbox{\tiny {gap}}}$ &&-6.67& &-7.2  & -6.53$^h$, -5.60$^k$ \\
InSb        &$b$       &&-1.88& &-2.0  & -2.3$^b$, -2.0$^{c,g}$,  -1.9$^f$  \\
\quad       &$d$       &&-4.62& &-4.7  & -4.8$^c$, -5.2$^b$ \\
\hline
\hline
\end{tabular}

\footnotesize $^a$Ref.~\cite{Meyer2001}, \footnotesize $^b$Ref.~\cite{FischettiJAP2010}, \footnotesize$^c$Ref.~\cite{Landolt1982}, 
\footnotesize $^d$Ref.~\cite{Williamson2000}, \footnotesize$^e$Ref.~\cite{Niquet2006}, \footnotesize $^f$Ref.~\cite{OReilly1992}, 
\footnotesize $^g$Ref.~\cite{Allan1988}, \footnotesize$^h$Ref.~\cite{Walle1989-1}, \footnotesize$^i$Ref.~\cite{Morgan1978}, 
\footnotesize$^j$Ref.~\cite{Tawinan2011}, \footnotesize$^k$Ref.~\cite{Wei-Zunger1999}.
\label{Table-deformation}
\end{table*}

\begin{figure}
\begin{center}
\includegraphics[width=0.8\textwidth]{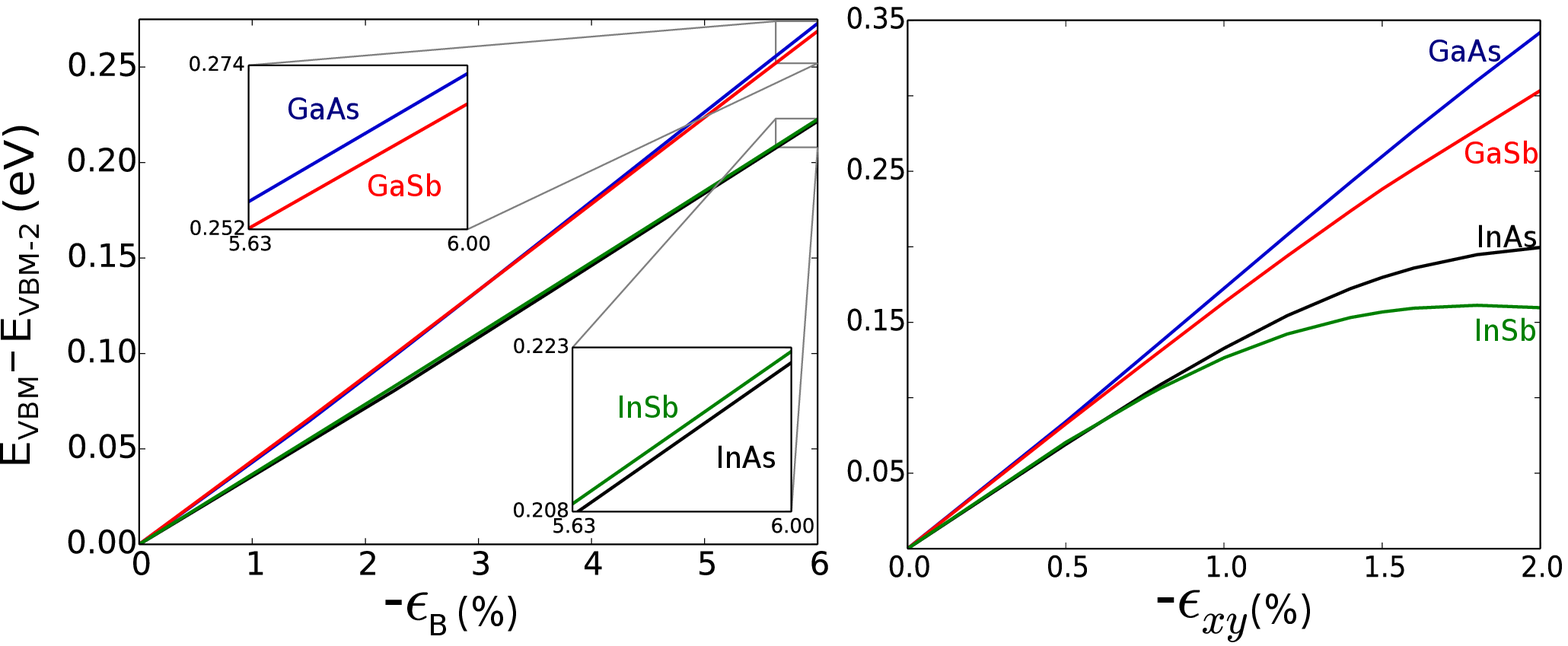}
\end{center}
\caption{Energy splitting between VBM and VBM-2 (which is the band just below VBM in the presence of spin degrees of freedom) 
obtained using hybrid functional DFT for negative $\epsilon_B \equiv \epsilon_{zz}-(\epsilon_{xx}+\epsilon_{yy})/2$ and negative 
$\epsilon_{xy}$ strains under uniaxial stress along [001] (left), and along [111] (right), respectively.}
\label{fig2}
\end{figure}

An interesting strain condition is when the valence band maximum (VBM) is no longer a pure $\left| \frac{3}{2},\frac{3}{2}\right\rangle$ 
state, namely a heavy hole band, but rather dominated by the light hole character of the $\left|\frac{3}{2},\frac{1}{2}\right\rangle$ state.
This occurs for the uniaxial stress applied along [001] and [111] directions provided that the biaxial strain $\epsilon_B \equiv \epsilon_{zz}-(\epsilon_{xx}+\epsilon_{yy})/2<0$, and $\epsilon_{xy}<0$ conditions hold, respectively.
Figure~\ref{fig2} illustrates the variation of the energy splitting between the two uppermost valence bands, VBM and VBM-2,
as a function of negative biaxial  $\epsilon_B$ and off-diagonal $\epsilon_{xy}$ strains under uniaxial stresses along [001] or [111] directions, 
respectively. A  cation-based grouping is clearly visible, that is, GaAs and GaSb behave similar, as do
InAs and InSb. Moreover, even though a linear trend is manifested under the [001] uniaxial stress for strains exceeding 6\%,
a nonlinear behavior can be observed at smaller strains in InAs and InSb under the [111] uniaxial stress. 
The nature of the uppermost valence bands will be further analyzed below in regard to the directional characteristics of the effective mass.

\begin{figure}
\begin{center}
\includegraphics[width=0.8\textwidth]{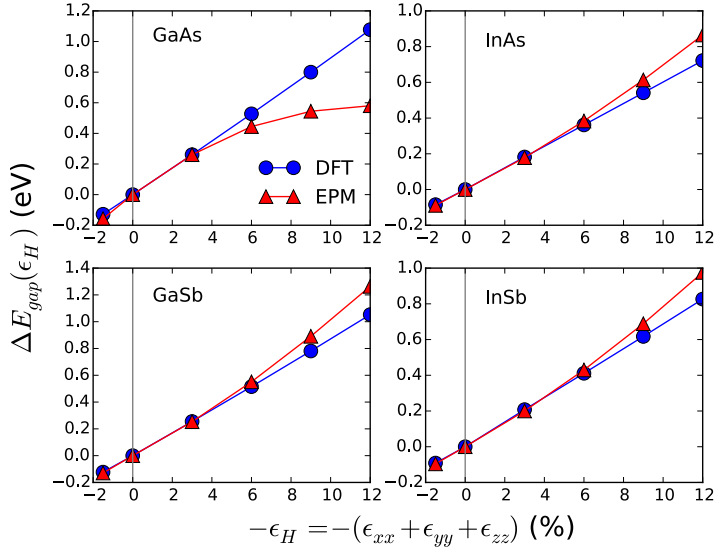}
\end{center}
\caption{Comparison of hybrid functional DFT versus EPM for the shift of direct bandgap from their unstrained values as a function of negative
hydrostatic strain.}
\label{fig3}
\end{figure}

\subsection{EPM Band Edge and Effective Mass Characteristics Under Strain}
We would like to contrast the band edge performance of EPM as described by the parameters contained in 
Table~\ref{Table-epm} with the hybrid DFT results in the presence of various strain conditions. 
First, we start with the hydrostatic strain, and compare in Figure~\ref{fig3} the shift of the direct bandgap from its unstrained value.
Since InSb and InAs become metallic, only a limited tensile strain is applied, whereas on the compressive side up to 12\% hydrostatic strain
is considered. The agreement of EPM with hybrid DFT results are seen to be excellent up to 6\% compressive strain, beyond which a deviation 
is observed, predominantly for GaAs. In practical considerations, as within this family GaAs has the smallest lattice constant and widest bandgap, it is generally used as the host matrix material, and therefore it seldom experiences such strain levels.

\begin{figure}
\begin{center}
\includegraphics[width=0.6\textwidth]{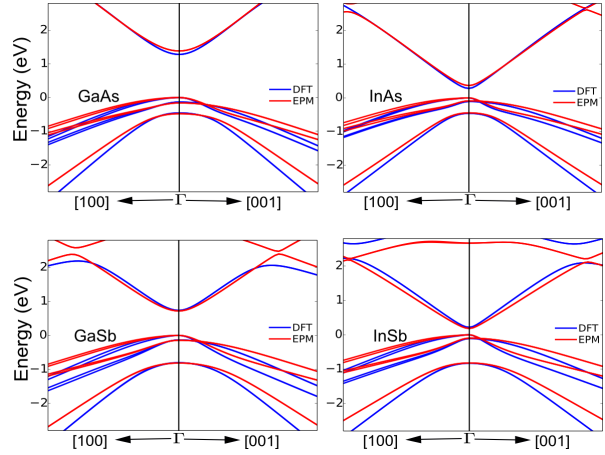}
\end{center}
\caption{Band edge characteristics of hybrid functional DFT and EPM under uniaxial stress along [001].
For each material VBM is set to zero energy level. The associated strain tensor is provided in the text.}
\label{fig4}
\end{figure}

\begin{figure}
\begin{center}
\includegraphics[width=0.6\textwidth]{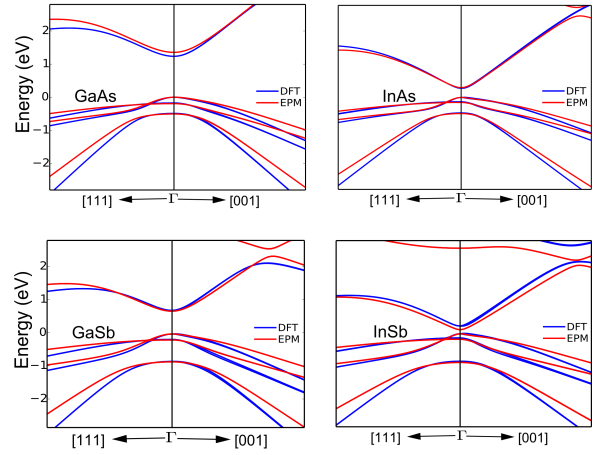}
\end{center}
\caption{Same as the previous figure, but under uniaxial stress along [111].}
\label{fig5}
\end{figure}

\begin{figure}
\begin{center}
\includegraphics[width=0.6\textwidth]{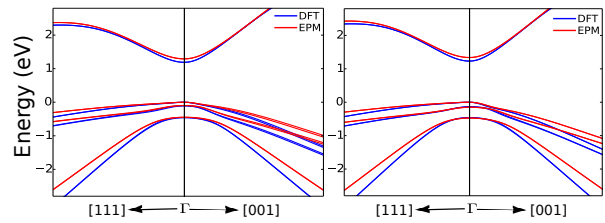}
\end{center}
\caption{Band edge characteristics  for GaAs of hybrid functional DFT and EPM for uniaxial stress along [120] (left), 
and for full anisotropic strain (right). The associated strain tensors are provided in the text.}
\label{fig6}
\end{figure}

Figures~\ref{fig4} and \ref{fig5} extend the comparison for all four materials to uniaxial stress along 
[001] and [111], respectively. For [001] stress, the strain tensor has the components $\epsilon_{xx}=\epsilon_{yy}=0.01,\, 
\epsilon_{zz}=-0.02$ with the other off-diagonal entries being zero, which corresponds to a biaxial strain of $\epsilon_{B}=-0.03$.
The [111] stress is reflected by the off-diagonal strain components of -0.01 and with all diagonal entries being zero.
As we fit EPM directly to the experimental values (cf. Table~\ref{Table-gaps}) a slight bandgap discrepancy is discernible 
for GaAs and InSb in these figures. Overall, it can be observed that the band edge behavior of EPM agrees well with the HSEsol results
within an energy span of at least a few hundred millielectronvolt of the respective band edge.
The reproduction of the crossings of the valence bands along the stress directions is particularly crucial.

In addition to these crystallographic directions of [001] and [111], we would like to test for uniaxial stress along [120]
direction as well as a full anisotropic strain governed by the tensors
\begin{eqnarray}
\epsilon_{\mbox{\tiny {[120]}}} & = & \left[
\begin{tabular}{ccc}
-0.001455 & 0.00842 & 0 \\
0.00842 & 0.02163 & 0 \\
0 & 0 & -0.00915 \\
\end{tabular}
\right]\, , \, \\
\epsilon_{\mbox{\tiny {full}}} & = & \left[
\begin{tabular}{ccc}
0.01 & 0.003 & 0.007 \\
0.003 & 0.01 & 0.02 \\
0.007 & 0.02 & -0.015 \\
\end{tabular}
\right]\, .
\end{eqnarray}
The comparison for these cases for GaAs shown in Figure~\ref{fig6} assures that the developed EPM fitting
faithfully represents HSEsol results under arbitrary strain profiles.
However, it needs to be noted that away from the band edges EPM starts to deviate from hybrid DFT,
as seen from Figures~\ref{fig4}, \ref{fig5} and \ref{fig6}. Thus, this set is specifically suitable 
for optical and excitonic characteristics around the $\Gamma$ point of the valence and conduction bands 
under diverse strain conditions.

\begin{figure}
\begin{center}
\includegraphics[width=0.6\textwidth]{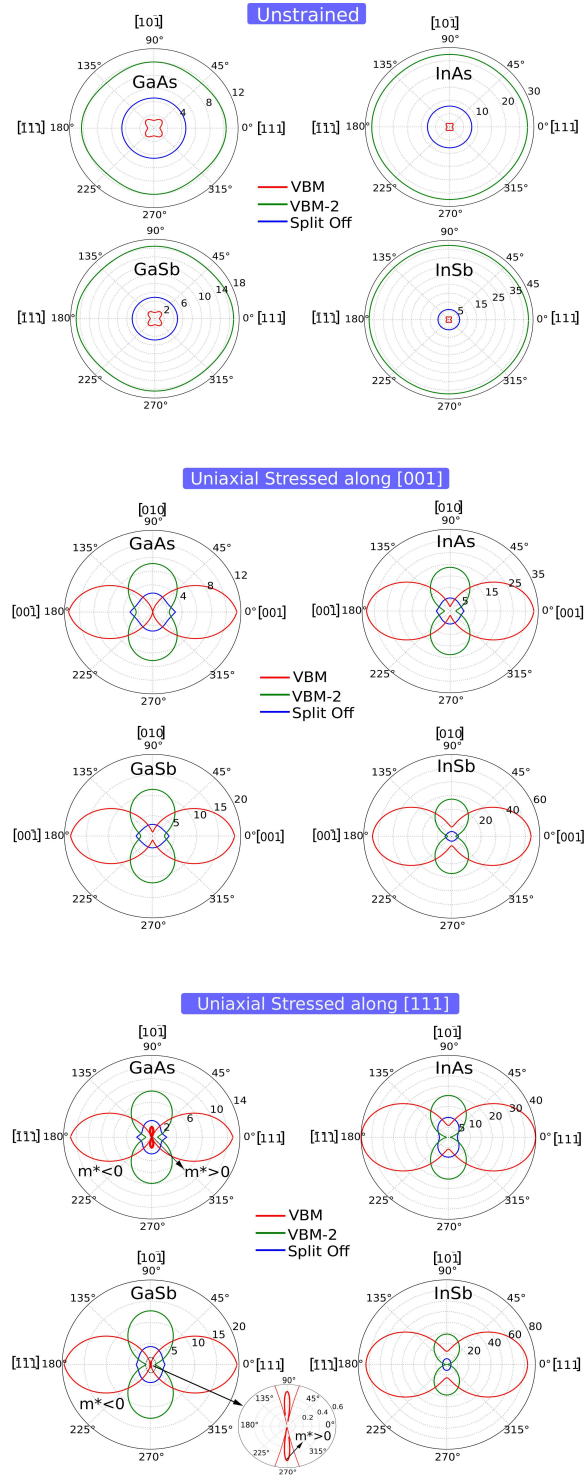}
\end{center}
\caption{Directional variation of EPM valence band inverse effective masses, $m_0/|m^*|$ for unstrained (top), and
uniaxially stressed along [001] (center), and [111] (bottom) crystal directions. Inset and thick red lines
highlight angular regions where VBM acquires electron-like positive effective mass.}
\label{fig7}
\end{figure}

Having established the band edge performance of the EPM set, we turn to the valence band 
effective mass characteristics in the absence and presence of strain. As we have seen in 
Figure~\ref{fig2}, deforming the crystal symmetry of the zinc-blende structure through stress 
removes the degeneracy at VBM. Now, we would like to explore the nature of the bands and in 
particular the directional mixing traits between the heavy and light hole 
bands with respect to the unstrained case. Figure~\ref{fig7} displays the directional variation 
of the effective mass of the top three valence bands as computed using EPM. Because of further band couplings 
under strain, the effective masses change sign, therefore we prefer to plot the 
\emph{inverses} of the effective masses. The top panel contains the unstrained case, where VBM is the heavy hole
band for all directions, and manifests the well-known warping behavior more distinctly than the 
light hole band \cite{hensel}.
The center and bottom panels of Figure~\ref{fig7} show the cases under uniaxial stress along [001] 
and [111] directions, respectively. 
The underlying strain tensors are the same as those for Figures~\ref{fig4} and \ref{fig5}, as quoted above.
For either case, the VBM exhibits light hole character along 
the stress direction, whereas in the perpendicular to stress direction the roles are swapped. 
As a matter of fact, for [111] uniaxial stress in GaAs and GaSb, along perpendicular to stress 
direction the VBM even changes sign and behaves like an electron band, see the inset at the bottom panel. 
The split-off band, being energetically remote 
from the upper two valence bands, preserves its isotropic behavior more or less in all the cases considered
above.
In general terms, our analysis supports recently measured higher hole mobility in group-III-V arsenides, and
group-III-V antimonides under strain \cite{nainai}, even though there is a non-trivial directional mixing
between heavy and light hole bands.

\section{Conclusions}
For the technologically important semiconductors of GaAs, GaSb, InAs and InSb, hybrid DFT 
calculations are employed to extract accurate band edge deformation potentials. 
Relying on this first-principles data, we offer a new EPM parametrization 
with superior performance under arbitrarily strained conditions. Through these EPM band structures, 
we demonstrate how the valence bands change character as a function of orientation under uniaxial stresses.
This reveals that VBM shows a light hole nature along the stress direction while displaying heavy hole
behaviour toward perpendicular directions.
Since this has implications in the optical selection rules and spin injection, it can be of importance for spintronics
or other quantum technologies \cite{zielinski13,huo2014}.
Given the reliability of our scheme, we believe that through a similar hybrid DFT study, the conduction band 
shear deformation potentials of the higher-lying degenerate valleys which are left outside the scope of this work 
can also be extracted.

\ack{We would like to thank T\"UB\.ITAK, The Scientific and Technological Research Council 
of Turkey for financial support through the project No. 112T178.
The numerical calculations reported in this paper were partially performed at T\"UB\.ITAK ULAKB\.IM, 
High Performance and Grid Computing Center (TRUBA resources).}

\section*{References}
\bibliography{manuscript}

\end{document}